\documentclass[journal]{IEEEtran}
\usepackage[latin9]{inputenc}
\usepackage{geometry}
\usepackage{color}
\usepackage{float}
\usepackage{amsmath}
\usepackage{amssymb}
\usepackage{graphicx}
\usepackage{empheq}
\makeatletter

\floatstyle{ruled}
\newfloat{algorithm}{tbp}{loa}
\providecommand{\algorithmname}{Algorithm}
\floatname{algorithm}{\protect\algorithmname}

\usepackage{amsfonts}
\usepackage{mathrsfs}
\usepackage{mathrsfs}\usepackage[noblocks]{authblk}
\usepackage[compress]{cite}
\usepackage{bm}
\usepackage{algorithm}
\usepackage{algorithmic}
\usepackage{epsfig}\usepackage{graphics}\usepackage{subfigure}\usepackage{epsfig}\usepackage{epstopdf}
\usepackage{graphics}\usepackage{subfigure}\usepackage{upgreek}\usepackage{caption}\usepackage{theorem}
\usepackage{breqn}
\usepackage{multicol}
\usepackage{eufrak}
\usepackage{eucal}
\usepackage{array}
\usepackage[compress]{cite}
\usepackage{algorithm}
\usepackage{algorithmic}
\theoremheaderfont{\normalfont\bfseries}

\makeatother

\begin{document}
\title{Secure Wireless Communication in RIS-Aided MISO System with Hardware
Impairments}
\author{Gui~Zhou, Cunhua~Pan, Hong~Ren, Kezhi~Wang, Zhangjie Peng% For a paper whose authors are all at the same institution,
 \thanks{(Corresponding author: Cunhua Pan) G. Zhou and C. Pan are with the
School of Electronic Engineering and Computer Science at Queen Mary
University of London, London E1 4NS, U.K. (e-mail: g.zhou, c.pan@qmul.ac.uk).
H. Ren is with the National Mobile Communications Research Laboratory,
Southeast University, Nanjing 210096, China. (hren@seu.edu.cn). K.
Wang is with Department of Computer and Information Sciences, Northumbria
University, UK. (e-mail: kezhi.wang@northumbria.ac.uk). Z. Peng is
with the College of Information, Mechanical and Electrical Engineering,
Shanghai Normal University, Shanghai 200234, China (e-mails: pengzhangjie@shnu.edu.cn).}\vspace{-0.3cm}
 }
\maketitle
\begin{abstract}
In practice, residual transceiver hardware impairments inevitably
lead to distortion noise which causes the performance loss. In this
paper, we study the robust transmission design for a reconfigurable
intelligent surface (RIS)-aided secure communication system in the
presence of transceiver hardware impairments. We aim for maximizing
the secrecy rate while ensuring the transmit power constraint on the
active beamforming at the base station and the unit-modulus constraint
on the passive beamforming at the RIS. To address this problem, we
adopt the alternate optimization method to iteratively optimize one
set of variables while keeping the other set fixed. Specifically,
the successive convex approximation (SCA) method is used to solve
the active beamforming optimization subproblem, while the passive
beamforming is obtained by using the semidefinite program (SDP) method.
Numerical results illustrate that the proposed transmission design
scheme is more robust to the hardware impairments than the conventional
non-robust scheme that ignores the impact of the hardware impairments. 
\end{abstract}


\section{Introduction}

Physical layer security issue has become a critical concern in current
and future wireless networks \cite{PLS-2015}. Fortunately, reconfigurable
intelligent surface (RIS) is promising for its application in  physical
layer security owing to the fact that the RIS can reconfigure electromagnetic
propagation environment for spectral and energy efficiency enhancement
in various applications \cite{Pan-jsac,Pan-twc,xianghao-2020,ning-letter,reviwer-1}.
In particular, an RIS, which is the key enabler to realize the vision
of smart radio environments \cite{Marco-6,Pan-mag}, is an inexpensive
adaptive (smart) thin panel that is composed of a number of nearly-passive
reflecting elements with controllable phase shifts. RIS is capable
of reconfiguring the reflected signal constructively for the signal
power enhancement at the legitimate users or destructively for avoiding
the information leakage to the eavesdroppers. The authors in \cite{reviwer-3} derived a closed-form solution to the secrecy rate maximization problem for an RIS-aided  single-user wireless communication system with one eavesdropper. The extension to the
	more general case with multiple eavesdroppers and multiple legitimate users
	was studied in \cite{reviwer-4}. Furthermore, Yu \textit{et.al.
	}\cite{xianghao-robust} investigated the robust transmission design
	for an RIS-aided multiuser secure communication system by considering
	the imperfect channel state information (CSI).  Then, an RIS-aided secure  system was studied  in \cite{hong-tcom}, where the artificial noise was used to enhance the security performance.

However, all the above-mentioned contributions on RIS-aided secure
communications assume the perfect hardware at the legitimate users,
which is too ideal in practice. The practical radio-frequency (RF)
components suffer from inevitable hardware impairments, including
phase noise, quantization errors, amplification noise, and nonlinearities
\cite{emil-HWI}. Both analytical and experimental results verify
that the distortion noise caused by the residual transceiver hardware
impairments can be modeled as additive Gaussian distribution, whose
variance is proportional to the signal power \cite{emil-HWI}. Recent
works have investigated the impact of hardware impairments on RIS-aided
systems \cite{HWI-3,HWI-1,HWI-2}. A closed-form expression for the
outage probability in an RIS-aided system was derived in \cite{HWI-3},
which revealed the fact that the hardware imperfection has prominent
impact on the achievable spectral efficiency. Robust beamforming design
was proposed in \cite{HWI-1} to maximize the received signal-to-noise
ratio (SNR) for an RIS-aided single-user multiple-input single-output
(MISO) system by considering the transceiver hardware impairments.
Furthermore, both uplink channel estimation and downlink transceiver
design were investigated in a single-user MISO system \cite{HWI-2}
by considering the RF impairments at the BS and phase noise at the
IRS.

To the best of our knowledge, the robust transmission design for an
RIS-aided secure wireless communication system in the presence of
transceiver hardware impairments has not been studied. To enhance
the secrecy rate of the system, we jointly design the active beamforming
at the BS and the passive beamforming at the RIS by considering the
hardware imperfection. The formulated optimization problem is solved
by resorting to alternating optimization, successive convex approximation,
and semidefinite relaxation. The simulation results demonstrate the
performance advantages of the proposed robust design scheme over the
benchmark schemes.

\noindent \textbf{Notations:} Most of the notations in this paper
are standard. The symbol $|\cdot|$ denotes the modulus of a complex
scalar.  $\mathrm{diag}(\mathbf{x})$ is a diagonal matrix with the
entries of $\mathbf{x}$ on its main diagonal. $\widetilde{\mathrm{diag}}({\bf X})$
is a diagonal matrix whose diagonal entries are the diagonal elements
of matrix ${\bf X}$. $\mathbf{X}\succeq\mathbf{Y}$
means that $\mathbf{X}-\mathbf{Y}$ is positive semidefinite. Additionally,
the symbol $\mathbb{C}$ denotes complex field and $\mathbb{R}$ represents
real field.

\section{System Model}

\begin{figure}
\centering \includegraphics[scale=0.6]{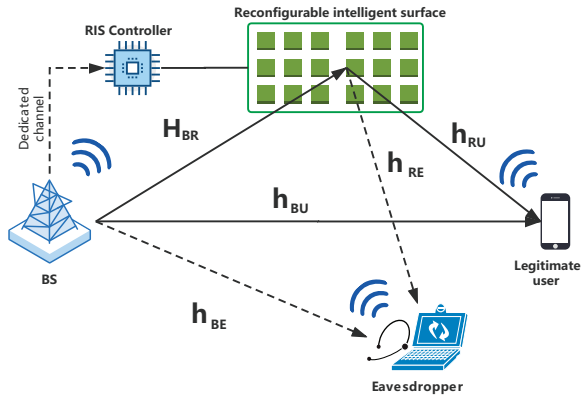} \caption{An IRS-aided MISO downlink secure communication system.}
\label{Figsysmodel} 
\end{figure}

\subsection{Signal Transmission Model}

Consider the downlink of an RIS-aided secure communication system
where a single-antenna eavesdropper intends to eavesdrop the confidential
signals sent by the BS to a single-antenna legitimate user, as shown
in Fig. \ref{Figsysmodel}\footnote{The proposed algorithm in this work for the single-user case can be directly applied to multi-user cases, including multiple legitimate users or multiple eavesdroppers}. The BS is equipped with $N$ active antennas.
An RIS equipped with $M$ programmable phase shifters is deployed
in the network to improve the physical layer security. Different from
the existing works, we consider the residual hardware impairments
at the BS and the legitimate user. Therefore, the signal transmitted
by the BS is given by 
\begin{equation}
\mathbf{x}={\bf f}s+\mathbf{m}_{t},
\end{equation}
where $s$ denotes the confidential independent Gaussian data symbol
and follows $\mathbb{E}[|s|^{2}]=1$, and ${\bf f}\in\mathbb{C}^{N\times1}$
is the corresponding beamforming vector. $\mathbf{m}_{t}\in\mathbb{C}^{N\times1}$
is the independent Gaussian transmit distortion noise and the power
of the distortion noise at each antenna is proportional to its transmit
signal power, i.e., $\mathbf{m}_{t}\sim\mathcal{C}\mathcal{N}(\mathbf{0},\mu_{t}\widetilde{\mathrm{diag}}({\bf f}{\bf f}^{\mathrm{H}}))$,
where $\mu_{t}\geq0$ is the ratio of transmit distorted noise power
to transmit signal power. This model has been widely used in the existing
literature \cite{emil-HWI,HWI-3,HWI-1,HWI-2} and is supported by
theoretical investigations and measurements \cite{emil-HWI} to accurately
model the joint effect of imperfect components in RF chains of practical
multiple-antenna systems.

The channels between the BS and the RIS, the BS and the legitimate
user, and the RIS and the legitimate user are denoted by $\mathbf{H}_{\mathrm{BR}}\in\mathbb{C}^{M\times N}$,
$\mathbf{h}_{\mathrm{BU}}\in\mathbb{C}^{N\times1}$ and $\mathbf{h}_{\mathrm{RU}}\in\mathbb{C}^{M\times1}$,
respectively. Then, the downlink received signal at the legitimate
user can be expressed as 
\begin{align}
y_{\mathrm{U}}  =\widetilde{y}_{\mathrm{U}}+m_{\mathrm{U}},
\end{align}
where  $\widetilde{y}_{\mathrm{U}}=(\mathbf{h}_{\mathrm{BU}}^{\mathrm{H}}+\mathbf{h}_{\mathrm{RU}}^{\mathrm{H}}\mathbf{E}\mathbf{H}_{\mathrm{BR}})\mathbf{x}+n_{\mathrm{U}}$. $n_{\mathrm{U}}\sim\mathcal{C}\mathcal{N}(0,\sigma_{\mathrm{U}}^{2})$
represents the additive white Gaussian noise (AWGN) at the legitimate
user with noise variance $\sigma_{\mathrm{U}}^{2}$. $\mathbf{E}=\mathrm{diag}(e_{1},...,e_{M})$, where $e_{m}$  is
the reflection coefficient on the $m$-th reflecting element of the
RIS. $m_{\mathrm{U}}$ is an independent zero-mean Gaussian distortion
noise at the legitimate user, which follows the distribution of $\mathcal{C}\mathcal{N}(0,\mu_{r}\mathrm{\mathbb{E}\{|\widetilde{y}_{\mathrm{U}}|^{2}\}})$,
where $\mu_{r}\geq0$ is the ratio of distorted noise power to undistorted
received signal power \cite{2012JSAC-HI}.

Similar to \cite{secrecy-HMI}, since the BS cannot acquire the full
knowledge about the eavesdropper, we consider the worst-case that
the eavesdropper has a high-quality hardware such that there is no
residual hardware impairments. Thus, the signal received by the eavesdropper
can be modeled as 
\begin{equation}
y_{\mathrm{E}}=(\mathbf{h}_{\mathrm{BE}}^{\mathrm{H}}+\mathbf{h}_{\mathrm{RE}}^{\mathrm{H}}\mathbf{E}\mathbf{H}_{\mathrm{BR}})\mathbf{x}+n_{\mathrm{E}},
\end{equation}
where $n_{\mathrm{E}}$ is the AWGN following the distribution of
$\mathcal{C}\mathcal{N}(0,\sigma_{\mathrm{E}}^{2})$. $\mathbf{h}_{\mathrm{BE}}\in\mathbb{C}^{N\times1}$
and $\mathbf{h}_{\mathrm{RE}}\in\mathbb{C}^{M\times1}$ are the channels
of the BS-eavesdropper and the RIS-eavesdropper links, respectively.
In this work, we assume that the BS perfectly knows all the CSI in
the entire network.

The secrecy rate of the system in can be expressed as 
\begin{equation}
R=[R_{\mathrm{U}}-R_{\mathrm{E}}]^{+},\label{eq:secure-rate}
\end{equation}
where $[a]^{+}\triangleq\max\left(a,0\right)$.

$R_{\mathrm{U}}$ is the rate of the legitimate user and is written
as 
\begin{align}
R_{\mathrm{U}} & =\log_{2}\left(1+\frac{1}{\Phi_{\mathrm{U}}(\mathbf{f},\boldsymbol{\mathbf{e}})}\mathbf{f}^{\mathrm{H}}\mathbf{G}_{\mathrm{U}}^{\mathrm{H}}\mathbf{e}\mathbf{e}^{\mathrm{H}}\mathbf{G}_{\mathrm{U}}{\bf f}\right)\label{rate-u}
\end{align}
where $\mathbf{e}=[e_{1},\cdots e_{M},1]^{\mathrm{H}}\in\mathbb{C}^{(M+1)\times1}$
is the equivalent reflection coefficient vector, $\mathbf{G}_{\mathrm{U}}=\left[\begin{array}{c}
\mathrm{diag}(\mathbf{h}_{\mathrm{RU}}^{\mathrm{H}})\mathbf{H}_{\mathrm{BR}}\\
\mathbf{h}_{\mathrm{BU}}^{\mathrm{H}}
\end{array}\right]\in\mathbb{C}^{(M+1)\times N}$ is the equivalent channel, and $\Phi_{\mathrm{U}}(\mathbf{f},\boldsymbol{\mathbf{e}})=\mathbf{e}^{\mathrm{H}}\mathbf{G}_{\mathrm{U}}(\mu_{r}{\bf f}\mathbf{f}^{\mathrm{H}}+(1+\mu_{r})\mu_{t}\widetilde{\mathrm{diag}}({\bf f}\mathbf{f}^{\mathrm{H}}))\mathbf{G}_{\mathrm{U}}^{\mathrm{H}}\mathbf{e}+(1+\mu_{r})\sigma_{\mathrm{U}}^{2}$
is the sum power of distortion noise and thermal noise.

$R_{\mathrm{E}}$ is the rate of the eavesdropper and is expressed
as 
\begin{align}
R_{\mathrm{E}} & =\log_{2}\left(1+\frac{1}{\Phi_{\mathrm{E}}(\mathbf{f},\boldsymbol{\mathbf{e}})}\mathbf{f}^{\mathrm{H}}\mathbf{G}_{\mathrm{E}}^{\mathrm{H}}\mathbf{e}\mathbf{e}^{\mathrm{H}}\mathbf{G}_{\mathrm{E}}{\bf f}\right)\label{rate-e}
\end{align}
where $\mathbf{G}_{\mathrm{E}}=\left[\begin{array}{c}
\mathrm{diag}(\mathbf{h}_{\mathrm{RE}}^{\mathrm{H}})\mathbf{H}_{\mathrm{BR}}\\
\mathbf{h}_{\mathrm{BE}}^{\mathrm{H}}
\end{array}\right]\in\mathbb{C}^{(M+1)\times N}$ is the equivalent channel, and $\Phi_{\mathrm{E}}(\mathbf{f},\boldsymbol{\mathbf{e}})=\mu_{t}\mathbf{f}^{\mathrm{H}}\widetilde{\mathrm{diag}}(\mathbf{G}_{\mathrm{E}}^{\mathrm{H}}\mathbf{e}\mathbf{e}^{\mathrm{H}}\mathbf{G}_{\mathrm{E}}){\bf f}+\sigma_{\mathrm{E}}^{2}$
is  the sum power of distortion noise and thermal noise.

\subsection{Problem Formulation}

In this work, we propose to maximize the secrecy rate $R$ given in
(\ref{eq:secure-rate}) by jointly optimizing the beamforming vector
at the BS and the passive beamforming at the RIS. The problem is formulated
as \begin{subequations} \label{Problem-1} 
\begin{align}
\max_{\mathbf{f},\boldsymbol{\mathbf{e}}} & \quad  R_{\mathrm{U}}(\mathbf{f},\boldsymbol{\mathbf{e}})-R_{\mathrm{E}}(\mathbf{f},\boldsymbol{\mathbf{e}})\label{obj11}\\
\mathrm{s}.\mathrm{t}. & \quad||\mathbf{f}||_{2}^{2}\leqslant P_{max},\label{eq:power-constraint}\\
 & \quad\boldsymbol{\mathbf{e}}\in\mathcal{S},\label{unit-modulus constraint}
\end{align}
\end{subequations} where (\ref{eq:power-constraint}) is the transmit
power constraint with the maximum transmit power $P_{max}$ at the
BS, and (\ref{unit-modulus constraint}) imposes a unit modulus on
each entry in $\boldsymbol{\mathbf{e}}$ with the set $\mathcal{S}=\{\mathbf{e}\mid|e_{m}|^{2}=1,1\leq m\leq M,e_{M+1}=1\}$. We remark that the optimal value of Problem (\ref{Problem-1}) is always non-negative. If $R_{\mathrm{U}}-R_{\mathrm{E}}<0$, we can set $||\mathbf{f}||_{2}^{2}=0$ such that the value of the objective function is equal to zero.

Problem (\ref{Problem-1}) is challenging to solve due to the non-concave
objective function and the non-convex constraint (\ref{unit-modulus constraint}).
Furthermore, in contrast to \cite{Shen2019secrecy} that a closed-form
solution was derived for $\mathbf{f}$, it is not straightforward
to obtain a closed-form solution for $\mathbf{f}$ in Problem (\ref{Problem-1})
due to the fact that the distortion noise introduced by the hardware
impairments aggregate the complexity of the secrecy rate expression.

\section{Algorithm design for secure wireless communications}

In this section, we adopt the alternating optimization (AO) method
to address the coupling of the beamforming vector at the BS and the
reflection beamforming at the RIS in Problem (\ref{Problem-1}). Specifically,
we alternately update $\mathbf{f}$ and $\mathbf{e}$ while fixing
the other variable.

First, due to (\ref{rate-u}) and (\ref{rate-e}), we introduce auxiliary variables $\mathbf{p}=[p_{1},p_{2},p_{3},p_{4}]^{\mathrm{T}}$, which satisfy
\begin{subequations}
	\label{cons-o}
	\begin{empheq}[left=\empheqlbrace]{align}
	&\log_{2}\left(\Phi_{\mathrm{U}}(\mathbf{f},\boldsymbol{\mathbf{e}})+\mathbf{f}^{\mathrm{H}}\mathbf{G}_{\mathrm{U}}^{\mathrm{H}}\mathbf{e}\mathbf{e}^{\mathrm{H}}\mathbf{G}_{\mathrm{U}}{\bf f}\right)\geq p_{1},\label{1o}  \\
	&\log_{2}\left(\Phi_{\mathrm{U}}(\mathbf{f},\boldsymbol{\mathbf{e}})\right)\leq p_{2},\label{2o}\\
	& \log_{2}\left(\Phi_{\mathrm{E}}(\mathbf{f},\boldsymbol{\mathbf{e}})+\mathbf{f}^{\mathrm{H}}\mathbf{G}_{\mathrm{E}}^{\mathrm{H}}\mathbf{e}\mathbf{e}^{\mathrm{H}}\mathbf{G}_{\mathrm{E}}{\bf f}\right)\leq p_{3},\label{3o}\\
	& \log\left(\Phi_{\mathrm{E}}(\mathbf{f},\boldsymbol{\mathbf{e}})\right)\geq p_{4},\label{4o} 
	\end{empheq}
	% \vspace{-3mm}
\end{subequations} such that  $R_{\mathrm{U}}\geq p_{1}-p_{2}$ and  $R_{\mathrm{E}}\leq p_{3}-p_{4}$. With (\ref{cons-o}), Problem (\ref{Problem-1}) is reformulated as\begin{subequations}
\label{Problem-3} 
\begin{align}
\max_{\mathbf{f},\mathbf{e},\mathbf{p}} & \quad p_{1}-p_{2}-p_{3}+p_{4}\\
\mathrm{s}.\mathrm{t}. & \quad (\ref{eq:power-constraint}),(\ref{unit-modulus constraint}),(\ref{cons-o}).
\end{align}
\end{subequations}

By using contradiction method, it can be readily proved that constraints in
(\ref{cons-o}) hold with equality at
the optimum. Hence, Problem (\ref{Problem-3}) is equivalent to Problem (\ref{Problem-1}).

\subsection{Optimize $\mathbf{f}$ with fixed $\mathbf{e}$}

When $\mathbf{e}$ is given, we further introduce auxiliary variables
$\mathbf{r}_{f}=[r_{f,1},r_{f,2},r_{f,3},r_{f,4}]^{\mathrm{T}}$ such
that non-convex constraints (\ref{cons-o})
are respectively equivalent to 
\begin{subequations}
	\label{cons-u}
	\begin{empheq}[left=(\ref{1o})(\ref{2o})\Rightarrow\empheqlbrace]{align}
		&\log_{2}(r_{f,1})\geq p_{1},\label{1f}  \\
		&\log_{2}(r_{f,2})\leq p_{2},\label{2f}\\
	    & \Phi_{\mathrm{U}}(\boldsymbol{\mathbf{f}})+\mathbf{f}^{\mathrm{H}}\mathbf{G}_{\mathrm{U}}^{\mathrm{H}}\mathbf{e}\mathbf{e}^{\mathrm{H}}\mathbf{G}_{\mathrm{U}}{\bf f}\geq r_{f,1},\label{3f}\\
        & \Phi_{\mathrm{U}}(\mathbf{f})\leq r_{f,2},\label{4f} 
	\end{empheq}
	% \vspace{-3mm}
\end{subequations}and
\begin{subequations}
	\label{cons-e}
	\begin{empheq}[left=(\ref{3o})(\ref{4o})\Rightarrow\empheqlbrace]{align}
		&\log_{2}(r_{f,3})\leq p_{3},\label{1e}\\
		&\log_{2}(r_{f,4})\geq p_{4},\label{2e}\\
		& \Phi_{\mathrm{E}}(\mathbf{f})+\mathbf{f}^{\mathrm{H}}\mathbf{G}_{\mathrm{E}}^{\mathrm{H}}\mathbf{e}\mathbf{e}^{\mathrm{H}}\mathbf{G}_{\mathrm{E}}{\bf f}\leq r_{f,3},\label{3e}\\
		&\Phi_{\mathrm{E}}(\mathbf{f})\geq r_{f,4}.\label{4e}
	\end{empheq}
	% \vspace{-3mm}
\end{subequations}It is observed from (\ref{cons-u}) and (\ref{cons-e})
that constraints (\ref{1f}), (\ref{4f}), (\ref{2e}), and (\ref{3e})
are convex, while constraints (\ref{2f}), (\ref{3f}), (\ref{1e}),
and (\ref{4e}) are concave. According to \cite{book-convex}, the
SCA can be used to address those concave constraints. In particular,
by adopting the first-order Taylor approximation and the equality
$\mathbf{y}^{\mathrm{H}}\widetilde{\mathrm{diag}}(\mathbf{x}\mathbf{x}^{\mathrm{H}})\mathbf{y}=\mathbf{x}^{\mathrm{H}}\widetilde{\mathrm{diag}}(\mathbf{y}\mathbf{y}^{\mathrm{H}})\mathbf{x}$,
we have the following equivalent constraints\begin{subequations}\label{12}
\begin{align}
(\ref{2f})\Rightarrow & \log_{2}(r_{f,2}^{n})+\frac{(r_{f,2}-r_{f,2}^{n})}{r_{f,2}^{n}\ln(2)}\leq p_{2},\label{eq:3}\\
(\ref{3f})\Rightarrow & 2\mathrm{Re}\left\{ \mathbf{f}^{n,\mathrm{H}}\mathbf{A}_{1}{\bf f}\right\} -\mathbf{f}^{n,\mathrm{H}}\mathbf{A}_{1}{\bf f}^{n}+(1+\mu_{r})\sigma_{\mathrm{U}}^{2}\geq r_{f,1},\label{eq:r}\\
(\ref{1e})\Rightarrow & \log_{2}(r_{f,3}^{n})+\frac{(r_{f,3}-r_{f,3}^{n})}{r_{f,3}^{n}\ln(2)}\leq p_{3},\label{eq:ee}\\
(\ref{4e})\Rightarrow & 2\mathrm{Re}\left\{ \mathbf{f}^{n,\mathrm{H}}\mathbf{A}_{2}{\bf f}\right\} -\mathbf{f}^{n,\mathrm{H}}\mathbf{A}_{2}{\bf f}^{n}+\sigma_{\mathrm{E}}^{2}\geq r_{f,4},\label{eq:11}
\end{align}
\end{subequations}where $\mathbf{A}_{1}=(1+\mu_{r})\mathbf{G}_{\mathrm{U}}^{\mathrm{H}}\mathbf{e}\mathbf{e}^{\mathrm{H}}\mathbf{G}_{\mathrm{U}}+(1+\mu_{r})\mu_{t}\widetilde{\mathrm{diag}}(\mathbf{G}_{\mathrm{U}}^{\mathrm{H}}\mathbf{e}\mathbf{e}^{\mathrm{H}}\mathbf{G}_{\mathrm{U}})$
and $\mathbf{A}_{2}=\mu_{t}\widetilde{\mathrm{diag}}(\mathbf{G}_{\mathrm{E}}^{\mathrm{H}}\mathbf{e}\mathbf{e}^{\mathrm{H}}\mathbf{G}_{\mathrm{E}})$.
$r_{f,2}^{n}$, $r_{f,3}^{n}$, and ${\bf f}^{n}$ are the solutions
obtained at the $n$-th iteration.

Finally, the subproblem for solving ${\bf f}$ is formulated as\begin{subequations}
\label{Problem-2} 
\begin{align}
\max_{\mathbf{f},\mathbf{p},\mathbf{r}_{f}} & \quad p_{1}-p_{2}-p_{3}+p_{4}\\
\mathrm{s}.\mathrm{t}. & \quad (\ref{eq:power-constraint}),(\ref{1f}),(\ref{4f}),(\ref{2e}),(\ref{3e}),(\ref{12}).\label{eq:const-u-1}
\end{align}
\end{subequations}Problem (\ref{Problem-2}) is an SOCP and can be
solved by using the CVX tool.

\subsection{Optimize $\mathbf{e}$ with fixed $\mathbf{f}$}

In order to tackle the non-convex unit-modulus constraint $\boldsymbol{\mathbf{e}}\in\mathcal{S}$,
we adopt the semidefinite relaxation (SDR) technique to update $\mathbf{e}$.
In particular, by defining a new variable $\widetilde{\mathbf{E}}=\mathbf{e}\mathbf{e}^{\mathrm{H}}$,
constraint $\boldsymbol{\mathbf{e}}\in\mathcal{S}$ is replaced by
$\{\widetilde{\mathbf{E}}\succeq\mathbf{0},\mathrm{rank}(\widetilde{\mathbf{E}})=1,\widetilde{\mathrm{diag}}(\widetilde{\mathbf{E}})=\mathbf{I}_{M+1}\}$.
Furthermore, with fixed ${\bf f}$ and new auxiliary variables $\mathbf{r}_{e}=[r_{e,1},r_{e,2},r_{e,3},r_{e,4}]^{\mathrm{T}}$,
non-convex constraints (\ref{1o}) and (\ref{2o}) are equivalent to 
\begin{subequations}
	\label{cons-u-1}
	\begin{empheq}[left=(\ref{1o})(\ref{2o})\Rightarrow\empheqlbrace]{align}
		&\log_{2}(r_{e,1})\geq p_{1},\label{1f-1}\\
		&\log_{2}(r_{e,2})\leq p_{2},\label{2f-1}\\
		& \mathrm{Tr}\left\{ \mathbf{B}_{1}\mathbf{E}\right\} +(1+\mu_{r})\sigma_{\mathrm{U}}^{2}\geq r_{e,1},\label{3f-1}\\
		& \mathrm{Tr}\left\{ \mathbf{B}_{2}\mathbf{E}\right\} +(1+\mu_{r})\sigma_{\mathrm{U}}^{2}\leq r_{e,2},\label{4f-1}
	\end{empheq}
	% \vspace{-3mm}
\end{subequations}
where $\mathbf{B}_{1}=\mathbf{B}_{2}+\mathbf{G}_{\mathrm{U}}{\bf f}\mathbf{f}^{\mathrm{H}}\mathbf{G}_{\mathrm{U}}^{\mathrm{H}}$
and $\mathbf{B}_{2}=\mu_{r}\mathbf{G}_{\mathrm{U}}{\bf f}\mathbf{f}^{\mathrm{H}}\mathbf{G}_{\mathrm{U}}^{\mathrm{H}}+(1+\mu_{r})\mu_{t}\mathbf{G}_{\mathrm{U}}\widetilde{\mathrm{diag}}({\bf f}\mathbf{f}^{\mathrm{H}})\mathbf{G}_{\mathrm{U}}^{\mathrm{H}}$.
The non-convex constraints (\ref{3o}) and (\ref{4o}) are equivalent to
\begin{subequations}
\label{cons-e-1}
	\begin{empheq}[left=(\ref{3o})(\ref{4o})\Rightarrow\empheqlbrace]{align}
		&\log_{2}(r_{e,3})\leq p_{3},\label{1e-1}\\
		&\log_{2}(r_{e,4})\geq p_{4},\label{2e-1}\\
		& \mathrm{Tr}\left\{ \mathbf{B}_{3}\mathbf{E}\right\} +\sigma_{\mathrm{E}}^{2}\leq r_{e,3},\label{3e-1}\\
		&\mathrm{Tr}\left\{ \mathbf{B}_{4}\mathbf{E}\right\} +\sigma_{\mathrm{E}}^{2}\geq r_{e,4},\label{4e-1}
	\end{empheq}
	% \vspace{-3mm}
\end{subequations}
where $\mathbf{B}_{3}=\mathbf{B}_{4}+\mathbf{G}_{\mathrm{E}}{\bf f}\mathbf{f}^{\mathrm{H}}\mathbf{G}_{\mathrm{E}}^{\mathrm{H}}$
and $\mathbf{B}_{4}=\mu_{t}\mathbf{G}_{\mathrm{E}}\mathrm{\widetilde{\mathrm{diag}}}({\bf f}\mathbf{f}^{\mathrm{H}})\mathbf{G}_{\mathrm{E}}^{\mathrm{H}}$.
In (\ref{cons-u-1}) and (\ref{cons-e-1}), the only non-convex constraints
(\ref{2f-1}) and (\ref{1e-1}) can be approximated by using the first-order
Taylor approximation as\begin{subequations} \label{16}
\begin{align}
(\ref{2f-1})\Rightarrow & \log_{2}(r_{e,2}^{n})+\frac{(r_{e,2}-r_{e,2}^{n})}{r_{e,2}^{n}\ln(2)}\leq p_{2},\label{eq:3-1}\\
(\ref{1e-1})\Rightarrow & \log_{2}(r_{e,3}^{n})+\frac{(r_{e,3}-r_{e,3}^{n})}{r_{e,3}^{n}\ln(2)}\leq p_{3},\label{eq:ee-1}
\end{align}
\end{subequations} where $r_{e,2}^{n}$ and $r_{e,3}^{n}$ are the
solutions obtained at the $n$-th iteration.

Finally, the relaxed subproblem of Problem (\ref{Problem-3}) is formulated
as\begin{subequations} \label{Problem-2-1} 
\begin{align}
\max_{\widetilde{\mathbf{E}},\mathbf{p},\mathbf{r}_{e}} & \quad p_{1}-p_{2}-p_{3}+p_{4}\\
\mathrm{s}.\mathrm{t}. & \quad\widetilde{\mathbf{E}}\succeq\mathbf{0},\mathrm{rank}(\widetilde{\mathbf{E}})=1,\widetilde{\mathrm{diag}}(\widetilde{\mathbf{E}})=\mathbf{I}_{M+1},\\
 & \quad(\ref{1f-1}),(\ref{3f-1}),(\ref{4f-1}),(\ref{2e-1})-(\ref{4e-1}),(16).\label{eq:const-u-1-1}
\end{align}
\end{subequations}Problem (\ref{Problem-2-1}) is further relaxed to a convex SDP by applying SDR method and then  solved by the CVX tools. The suboptimal solution $\mathbf{v}$,
which maximizes the objective value of Problem (\ref{Problem-1}),
can be obtained from the optimal $\widetilde{\mathbf{E}}$ by using
Gaussian randomization techniques. Then, $\mathbf{v}$ is projected
onto the constraint set $\mathcal{S}$ as $\widetilde{\mathbf{e}}=\mathrm{exp}(\mathrm{j}\angle(\frac{\mathbf{v}}{[\mathbf{v}]_{M+1}}))$,
where $\mathrm{j}\triangleq\sqrt{-1}$, $\angle(x)$ denotes the angle
of $x$, and $[\mathbf{x}]_{m}$ denotes the $m$-th entry of vector
$\mathbf{x}$. Please note that the suboptimal solution $\widetilde{\mathbf{e}}$
generated by the SDR method cannot guarantee that the objective value
of Problem (\ref{Problem-1}) in the $(n+1)$-th iteration is no smaller
than that in the previous iteration. Therefore, in order to ensure
the non-decreasing objective value sequence generated in each iteration,
we adopt the following update 
\begin{equation}
\mathbf{e}^{n+1}=\begin{cases}
\widetilde{\mathbf{e}}, & \textrm{if }R\left(\mathbf{f}^{n+1},\widetilde{\mathbf{e}}\right)\geq R\left(\mathbf{f}^{n+1},\mathbf{e}^{n}\right),\\
\mathbf{e}^{n} & \textrm{otherwise}.
\end{cases}\label{eq:e-up}
\end{equation}
Note that when $\mathbf{e}^{n+1}=\mathbf{e}^{n}$, we still have $R\left(\mathbf{f}^{n+2},\mathbf{e}^{n+1}\right)>R\left(\mathbf{f}^{n+1},\mathbf{e}^{n+1}\right)=R\left(\mathbf{f}^{n+1},\mathbf{e}^{n}\right)$
and the iteration process does not terminate until convergence. The
reason is that the optimal $\mathbf{f}^{n+2}$ of the next iteration
is obtained based on $\mathbf{f}^{n+1}$ and $\mathbf{e}^{n}$.

\subsection{Algorithm Analysis}

\paragraph{Convergence analysis}

Firstly, we state that the objective value sequence $\{R(\mathbf{F}^{n},\mathbf{e}^{n})\}$
generated in each iteration in the AO algorithm is non-decreasing. In particular, it follows
that 
\begin{align*}
 & R\left(\mathbf{f}^{n+1},\mathbf{e}^{n+1}\right)\overset{\mathrm{(a)}}{\geq}R\left(\mathbf{f}^{n+1},\mathbf{e}^{n}\right)\overset{\mathrm{(b)}}{>}R\left(\mathbf{f}^{n},\mathbf{e}^{n}\right).
\end{align*}
The above (a) is due to the update in (\ref{eq:e-up}), and (b) follows
from the globally optimal solution $\mathbf{f}^{n+1}$ of Problem
(\ref{Problem-2}) for a given $\mathbf{e}^{n}$.
Moreover,  $R(\mathbf{F}^{n},\mathbf{e}^{n})$
has a finite upper bound due to the bounded transmit power constraint.
Therefore, the AO algorithm  is guaranteed to converge.

\paragraph{Complexity analysis}
According to \cite{Ben-Tal2001convex}, the complexity of solving the SOCP of Problem (\ref{Problem-2}) is  $\mathcal{O}(N^{3})$, and the complexity of solving the SDP corresponding to Problem (\ref{Problem-2-1}) is  $\mathcal{O}(M^{3.5})$.

\section{Numerical results and discussions}

In this section, the numerical results are provided to evaluate the
performance of the proposed algorithm for an RIS-aided system with
hardware impairments. The BS and the RIS are located at (0 m, 0 m)
and (50 m, 0 m), respectively. Both legitimate user and eavesdropper
lie in a horizontal line which is parallel to the BS-RIS path, which
means the coordinates of legitimate user and eavesdropper are (50
m, 2 m) and (45 m, 2 m), respectively. We set $N=4$ and $\sigma_{\mathrm{U}}^{2}=\sigma_{\mathrm{E}}^{2}=-80$
dBm. The large-scale path loss is $\mathrm{PL}=-30-10\alpha\log_{10}(d)$
dB, where $\alpha$ is the path loss exponent and $d$ is the link
length in meters. The path loss exponent of the RIS related channels
is set to $2.2$ and that of the direct BS-user and BS-eavesdropper
channels is set to $3.6$. The small-scale fading follows a Rician
distribution with a Ricean factor of 10 for the RIS related channels
and a Ricean factor of 0 for the direct channels. The line-of-sight
(LoS) components are defined by the product of the steering vectors
of the transmitter and receiver and the non-LoS components are drawn
from a Rayleigh fading. The proposed scheme and the benchmark schemes
are listed as: 1) ``RIS-robust'', the proposed robust beamforming
design compensating for the hardware impairments in an RIS-aided system;
2) ``NonRIS-robust'', a robust beamforming design compensating for
the hardware impairments in a conventional system without RIS; 3)
``RIS-nonrobust'', naive beamforming design ignoring the hardware
impairments in an RIS-aided system; 4) ``NonRIS-nonrobust'', naive
beamforming design ignoring the hardware impairments in a conventional
system without RIS.

\begin{figure}
\centering \includegraphics[width=3.4in,height=2.5in]{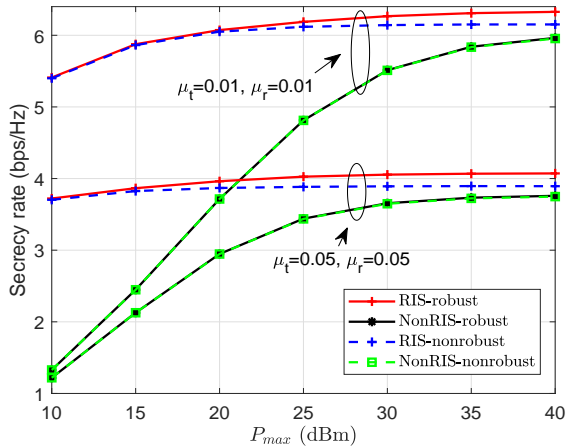}
\caption{Comparison of secrecy rate as a function of $P_{max},$when
$N=4$ and $M=32$.}
\label{rate}
\end{figure}

Fig. \ref{rate} investigates the secrecy rate of different schemes
as a function of the maximum transmit power at the BS when
$M=32$. Fig. \ref{rate} shows that the secrecy rate tends to be
stable with the increase of $P_{max}$. This is because the distortion
noise caused by hardware impairments is proportional to the transceiver
signal power, so there is an upper bound for the secrecy rate as a
function of the maximum transmit power. It is interesting to find
that the robust beamforming design in traditional systems without
RIS is not effective in compensating for the performance loss caused
by the hardware impairments. However, the robust design is more effective
in RIS-aided secure systems. These observations reveal the importance
of the robust design for RIS-aided secure systems by considering the
hardware impact.

\begin{figure}
\centering \includegraphics[width=3.4in,height=2.5in]{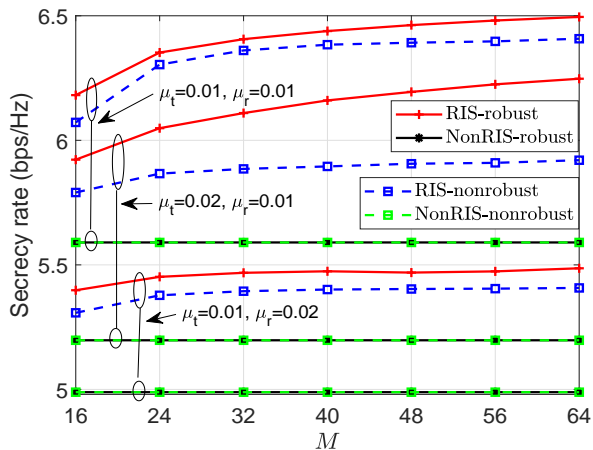} \caption{Comparison of secrecy rate as a function of $M,$when $N=4$ and $P_{max}=30$
dBm.}
\label{M} 
\end{figure}

Fig. \ref{M} illustrates the impact of the number of the reflecting
elements on the system performance when $P_{max}=30$ dBm. Firstly,
the increase of RIS's size can enhance the secrecy rate, but this
performance gain decreases when the level of the hardware impairments
(i.e., $\mu_{t}$ and $\mu_{r}$) increases. Furthermore, by comparing
the systems with $\{\mu_{t}=0.01,\mu_{r}=0.02\}$ and the systems
with $\{\mu_{t}=0.02,\mu_{r}=0.01\}$, the hardware impairments on
legitimate user has a greater negative impact on the secrecy rate
than that of the hardware impairments on the BS. This is because the
proportion of the receive distortion noise power in total received
noise is much larger than the transmit distortion noise power.

\section{Conclusions}

In this work, we have improved the performance of the secrecy communication
system in the presence of transceiver hardware impairments by employing
an RIS and designing robust beamforming. The secrecy rate of the system
was maximized by the joint optimization of the active beamforming
vector at the BS and the passive beamforming at the RIS when the distortion
noise caused by the hardware impairments was taken into account. The
two variables were updated alternately by solving two approximate
subproblems as the form of SOCP and SDP, respectively. Our simulation
results have demonstrated the performance advantage of the proposed
beamforming design.

 \bibliographystyle{IEEEtran}
\bibliography{bibfile}

\end{document}